\documentclass{webofc}
\usepackage[varg]{txfonts}  

\begin{document}

\title{Elucidating the $\rho$-meson's role as intermediate resonance 
in the time-like electromagnetic pion form factor}

\author{\firstname{Reinhard} \lastname{Alkofer}\inst{1}\fnsep\thanks{\email{reinhard.alkofer@uni-graz.at}} \and
        \firstname{\'Angel S.}~\lastname{Miramontes}\inst{2}\fnsep\thanks{\email{angel-aml@hotmail.com}} \and
        \firstname{H\`elios} \lastname{Sanchis Alepuz}\inst{3}\fnsep\thanks{\email{helios.sanchis-alepuz@silicon-austria.com}}
}

\institute{Institute of Physics, University of Graz, NAWI Graz, Universit\"atsplatz 5, 8010 Graz, Austria
\and
           Instituto de F\'isica y Matem\'aticas, Universidad Michoacana de San Nicol\'as de Hidalgo, 
\\ Morelia, Michoac\'an 58040, Mexico 
\and
           Silicon Austria Labs GmbH, Inffeldgasse 25F, 8010 Graz, Austria
          }

\abstract{Motivated by the planned measurements of the time-like electromagnetic proton form factors
 an  exploratory study of the
time-like electromagnetic pion form factor
is presented in a formalism which describes mesons as Poincar\'e-invariant bound states. 
In the respective quark interaction kernel, beyond the gluon-intermediated interactions for 
valence-type quarks, non-valence effects are included by allowing the pions to couple back 
in a self-consistent manner to the quarks. 
Consequently, the opening of the dominant $\rho$ decay channel,
$\rho \to \pi\pi$, and the presence of a multi-particle branch cut, setting in when the two-pion threshold is crossed, are included consistently, first in the correspondingly calculated 
quark-photon vertex, and then consequently in the time-like electromagnetic pion form factor.
The obtained results are in agreement with the available experimental data and 
provide further evidence for the efficacy of a vector-meson dominance model.
Effects of going beyond the here employed isospin-symmetric limit
are discussed. Last but not least, an outlook is provided on how to include 
in Poincar\'e-covariant Faddeev approaches
the effects of intermediate resonances in the calculation of baryon properties.
}
\maketitle
\section{Introduction}
\label{intro}

Quantum Chromodynamics (QCD) provides the basis for our current understanding of
hadron physics. Perturbative as well as non-perturbative QCD calculations 
have become impressively successful in the last decades. From a theoretical point of view,
QCD is a model Quantum Field Theory. It realises locality and unitarity, and being 
asymptotically free QCD is an ultraviolet complete Quantum Field Theory. The latter property
quite obviously underlies the success of perturbative calculations for hard hadronic 
and semi-hadronic processes.

The perturbative description of elementary particles is essentially based on the 
field-particle duality which means that each field in a Quantum Field Theory is associated 
with a physical particle. Quite obviously, this is not the case for the elementary fields of QCD, 
the quarks and gluons. First of all, QCD being a gauge theory, together with the fact that 
only gauge-invariant quantities can be observables, rules already out that quarks and gluons
can be directly observed. In a gauge theory, the r\^ole of fields is to implement the principle 
of locality. The types of different elementary fields needed in the theory is related to the 
structure of the gauge group, it is not, at least not directly, related to the empirical spectrum 
of particles. Second, what is usually referred to as ``confinement'' is obviously going beyond
the mere issue of ``only-gauge-invariant-observables'', respectively, colour confinement
(see, {\it e.g.}, refs.\ \cite{Greensite:2011zz,Alkofer:2006fu}), 
and an especially puzzling aspect is the following:
As only hadrons are produced from processes involving hadronic initial
states, one has to explain that the only thresholds in hadronic amplitudes
are due to the productions of other hadronic states. Consequently, singularities occurring 
in the amplitudes of hadronic and thus composite states are exclusively due to their 
hadronic substructure, or phrased otherwise, due to virtual fluctuations to hadronic resonances.   

This implies not only that hadron spectroscopy and hadron structure are strongly interrelated, 
but, in particular, that a microscopic understanding of the effect of resonances on form factors, 
structure functions, etc., will shed light on the precise working of quark-hadron duality.
Compelling evidence for the consequences of this duality on the qualitative as well as the 
semi-quantitative level has been found (for a review see, 
{\it e.g.}, Ref.\ \cite{Melnitchouk:2005zr}) and contributed to a verification of this duality 
which is, beyond the trivial fact of the absence of coloured states, the clearest experimental 
signature for confinement. The attributed perfect orthogonality of the quark and gluon
degrees of freedom  on the one hand and hadronic states on the other hand, and thus 
the perfect absence of ‘‘double-counting’’ when doing calculations either with quarks and gluons
within QCD or within purely hadronic descriptions, is nothing else but another way to express 
central aspects of confinement. 
Basing a (confessedly impossible) exact calculation for a hadronic or semi-hadronic 
scattering process either 
on elementary QCD correlation functions or only on hadronic correlations functions needs 
to lead to the same result by the virtue of colour confinement, the latter being a consequence
of gauge invariance. The way 
by which the hadronic amplitudes become effectively independent of the singularities of the
elementary correlation functions is then an imprint of the dynamical realisation of confinement.

The formation of an excited hadronic bound state, being an unstable resonance, is closely 
related to the open decay channels. This is most evident from the analysis of time-like form 
factors in kinematic regions close to a resonance. To capture this interplay correctly in a 
QCD-based calculation of a hadron  form factor, the in- (respectively, out-) going hadron  
and the hadronic resonance which is apparent in the form factor have to be described both
in a mutual consistent manner as composite objects of quarks and gluons. Thus, a calculation of 
a time-like form factor from QCD, or from a microscopic model based on QCD degrees of 
freedom, has to overcome the challenge to treat all necessary elements on the same basis
and to a sufficient degree of sophistication. Only then the result will allow conclusions on the 
hadron structure encoded in such a form factor.

It should be emphasised that the presented study aims for comprehending the interplay 
of the different features of the form factor and as they arise from the QCD degrees of 
freedom.  It is also designed  such that it can be generalised to an investigation of baryon
 form factors, and hereby especially the nucleons' time-like form factor. 
(NB: Different space-like baryon form factors have been calculated in this approach, 
see, {\it e.g.}, Refs.\
\cite{Eichmann:2016yit,Nicmorus:2010sd,Eichmann:2011vu,Eichmann:2011pv,Sanchis-Alepuz:2013iia,Sanchis-Alepuz:2017mir}
for some recent respective work.) 
A thorough understanding of the proton time-like form factor at very low $Q^2$ is 
a very timely subject as the upcoming PANDA experiment possesses the unique possibility to 
measure the proton's
electromagnetic form factors in the so-called unphysical region through the process 
$\bar p p \to l^+l^-\pi^0$, $l=e, \mu$
\cite{Fischer:2021kcr}. At large $Q^2$ the question of the onset of the convergence 
scale between the space-like and the time-like form factors arises.

In this talk an exploratory study of the time-like pion electromagnetic form factor 
employing a combination of Bethe-Salpeter and Dyson-Schwinger equations
 \cite{Miramontes:2021xgn} will be discussed, and its results will be presented. 
To obtain the most important features of the pions' time-like form factor one needs to take into
account\\
(i) the pion as bound state of quark and antiquark thereby at the same accommodating 
its special role as would-be Goldstone boson of the dynamically broken chiral symmetry of QCD,\\
(ii) the mixing of the $\rho$-meson, being determined consistently as a quark-antiquark bound
state, with a virtual photon when this photon is in turn coupled to a quark-antiquark pair via the 
fully renormalised quark-photon vertex, and\\
(iii) the dominant decay channel of the $\rho$-meson, namely, $\rho \to \pi \pi$. \\

Before presenting how to combine these in our study a few general remarks about
the time-like pion electromagnetic form factor are in order.

\section{\label{sec:PionFF} The time-like electromagnetic pion form factor: General aspects }

\begin{figure}[ht]
\caption{Electron-positron pair annihilating to a virtual photon with time-like momentum which then
decays to a pion pair. }
\label{Fig:PiFF}
\begin{center}
\setlength{\unitlength}{0.5mm}
\begin{picture}(150,60)
\thicklines
\multiput(32.5,30)(10,0){6}{\oval(5,5)[t]}
\multiput(37.5,30)(10,0){6}{\oval(5,5)[b]}
\put(60,40){$\gamma^\star $}
\put(10,50){\vector(1,-1){10}}
\put(20,40){\line(1,-1){10}}
\put(20,45){\Large $e^-$}
\put(30,30){\vector(-1,-1){10}}
\put(20,20){\line(-1,-1){10}}
\put(20,15){\Large $e^+$}
\put(100,30){\circle{20}}
\put(95,28.5){\Large $F_\pi$}
\put(125,45){\Large $\pi^+$}
\put(125,15){\Large $\pi^-$}
\multiput(108,35)(10,10){2}{\line(1,1){8}}
\multiput(108,25)(10,-10){2}{\line(1,-1){8}}
\end{picture}
\end{center}
\end{figure}
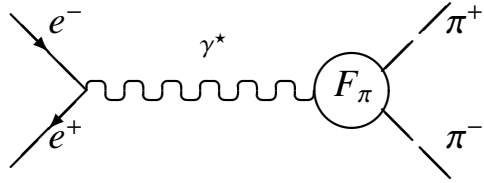

Experimentally the time-like electromagnetic pion form factor is typically determined from 
electron-positron pair annihilation into a pion pair, see Fig.~\ref{Fig:PiFF}. In the physical region
above the two-pion production threshold the  corresponding amplitude possesses a cut, and thus
the time-like pion form factor has to be treated as a complex quantity.

The pion form factor contains all kind of processes turning a virtual photon into a pion pair,
{\it cf.\/} Fig.\  \ref{Fig:PiFF}. Using that in the interesting kinematic regime strong-interaction
processes dominate by orders of magnitude against electroweak processes, an important first
ingredient to the pion form factor is how the virtual photon couples to a quark via all possible
QCD processes. The corresponding amplitude is nothing else than the full quark-photon 
vertex which includes, at least in principle, the information about the virtual photon's 
hadronic substructure. The two more further ingredients are the quark propagator 
and the pion bound state amplitude.

\begin{figure}[ht]
\centering
\sidecaption
\includegraphics[width=0.7\textwidth]{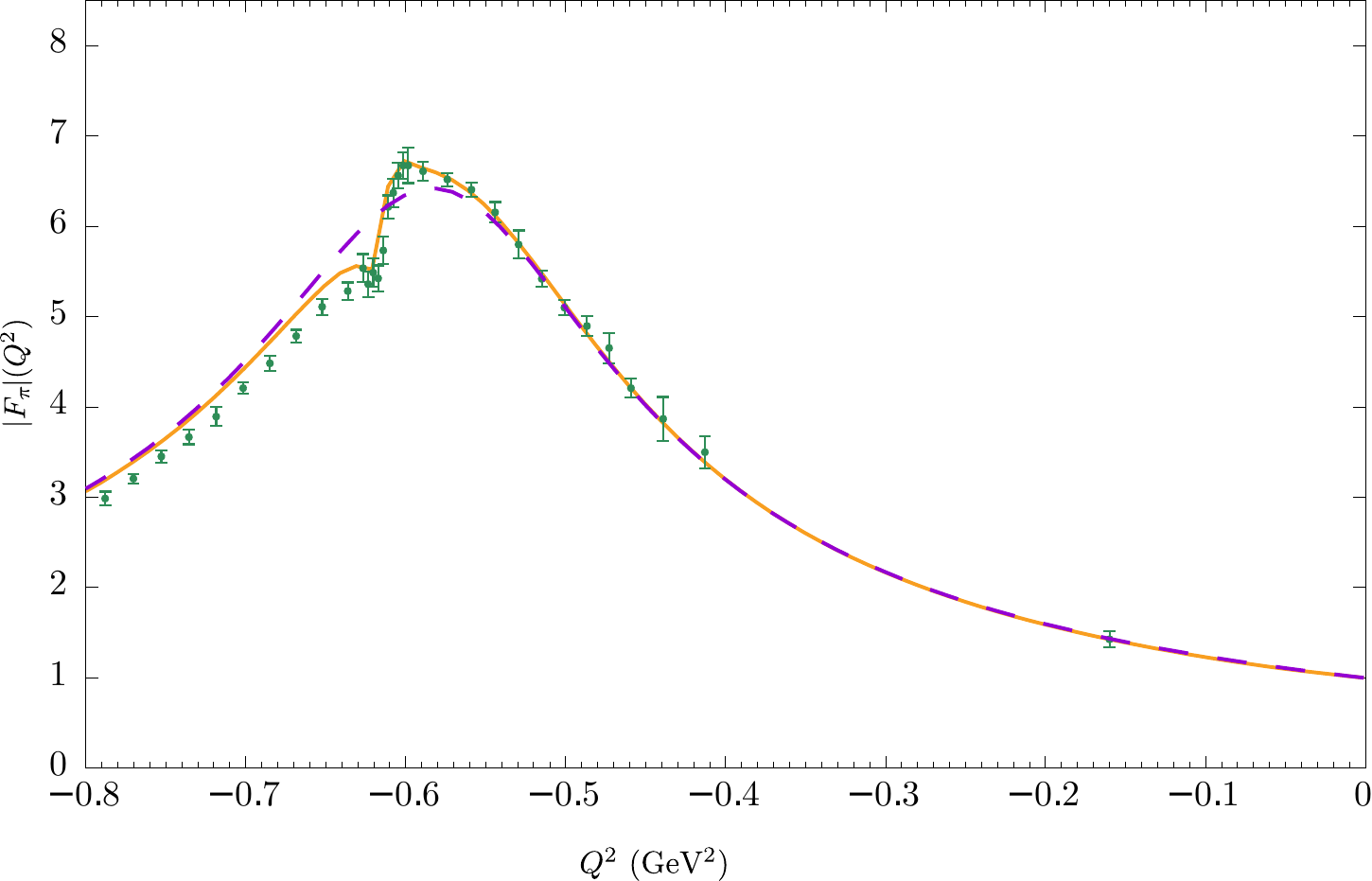}
\caption{Absolute value of the pion form factor in the time-like ($Q^2<0$) domain 
from the VMD-based
fit given in ref.\ \cite{OConnell:1995nse} (full line) in comparison to the experimental data
 \cite{Akhmetshin:2006bx}.
The dashed line is based on the same expression
but without $\rho$-$\omega$ mixing, {\it i.e.},  the mixing matrix element is put to zero. }
\label{fig:FpiVMDplot}   
\end{figure}

The presented study is performed in the isospin symmetric limit.\footnote{Isospin 
breaking and thus the effect of $\rho$-$\omega$ mixing on the pion form
factor in the presented approach is currently under investigation. 
Corresponding first results on electromagnetic and
strong isospin breaking effects on the mass splittings of light pseudo-scalar and 
vector mesons calculated from Bethe-Salpeter and Dyson-Schwinger equations
became recently available \cite{Miramontes:2022mex}.} 
As isospin 
breaking is clearly visible in the time-like pion form factor, in particular through
 $\rho$-$\omega$ mixing, a vector meson dominance (VMD) based fit is used 
to extract the expected form of the pion form factor without this mixing effect,
see Fig.\ \ref{fig:FpiVMDplot}.

As can be seen from Fig.\ \ref{fig:FpiVMDplot} the VMD model of ref.\ \cite{OConnell:1995nse}
describe the experimental data precisely. But how does it arise from the dynamics of QCD degrees
of freedom?

\section{Interactions in Dyson-Schwinger and Bethe-Salpeter equations}

\begin{figure*}[ht]
\caption{Truncations used herein for the Bethe-Salpeter interaction kernel $K$ (upper diagram) and 
the quark Dyson-Schwinger equation (lower diagram).
In the upper diagram, the terms on the right-hand side correspond to the 
gluon, pion exchange, and s- and u-channel pion decay contributions to the truncation, respectively. The s- and u-channel pion decay terms do not contribute to the quark Dyson-Schwinger equation.}
\label{fig:kernels}
\centering
\includegraphics[width=0.99\textwidth]{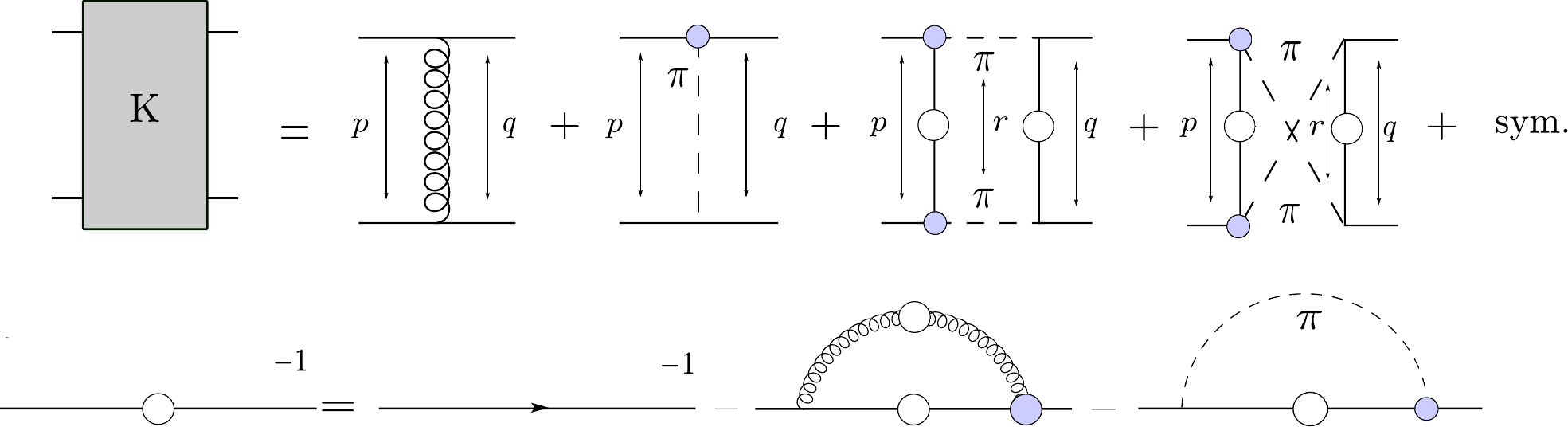}
\end{figure*}

The interactions used in the exploratory  study \cite{Miramontes:2021xgn} have been a 
gluon exchange (in the form of the Maris-Tandy model \cite{Maris:1997tm,Maris:1999nt})
as well as 
a pion exchange and $s$- and $u$-channel pion decay contributions, 
see Fig.\ \ref{fig:kernels}, the latter both in self-consistent manner. Here the following 
disclaimer is in order: To keep this calculation feasible a number of technically motivated approximations have been made. It has been checked that these had only a minor impact 
on the results, for further details see Ref.\ \cite{Miramontes:2021xgn}.

As already stated the quark-photon vertex is a decisive element in the presented calculation.
It has been determined with the above mentioned interactions in Ref.\ \cite{Miramontes:2019mco}.
The Dirac structure of the vertex can be expanded in a basis consisting of twelve elements, 
and all of them are taken into account to obtain the below displayed results. In addition, 
out of these twelve structure eight are transverse to the photon's momentum. These transverse
amplitudes display a pole at the (calculated) $\rho$-meson mass and multi-pion cut above the
two-pion threshold.\footnote{This quark-photon vertex has been recently used in a calculation
of the pion and the kaon box contribution to $a_{\mu}^{\text{HLbL}}$ \cite{Miramontes:2021exi}
demonstrating its usability also in the context of hadronic contributions to light-light scattering.}

\goodbreak

\begin{figure}[ht]
\centering
\sidecaption
\includegraphics[width=0.7\textwidth]{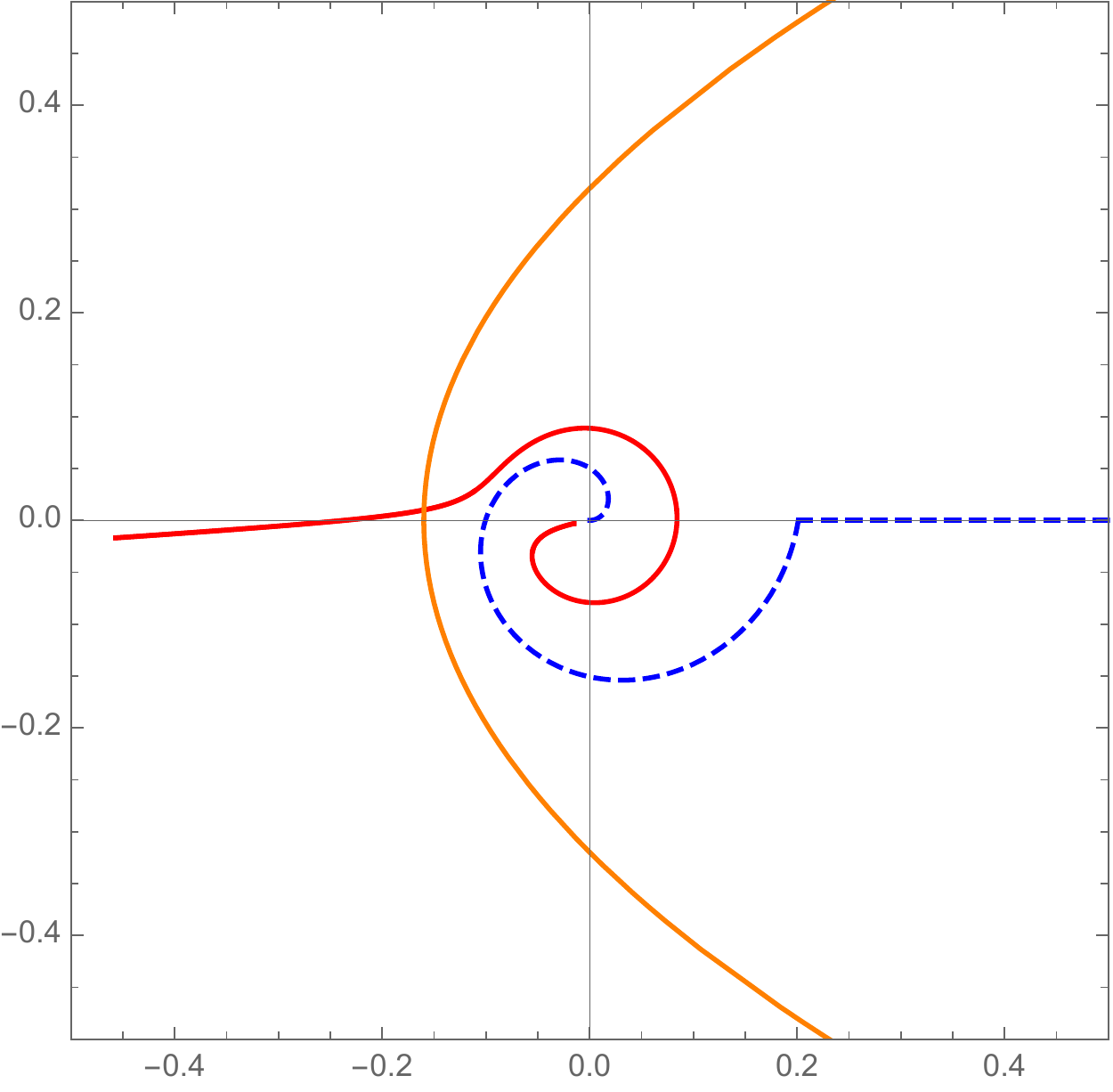}
\caption{Shown is the complex $p^2$ plane such that time-like momenta correspond to the 
left real half-axis. Without singularities in the integrand the integration would be performed 
over space-like momenta, {\it i.e.}, over the right real half-axis. In the presented example 
this is not possible due to the pion cut represented by the red fill line. Correspondingly, the 
integration contour is deformed to the dashed blue line. The singularities of the quark propagator
are all to the left of the displayed parabola, and therefore do not necessitate further 
contour deformations. }
\label{fig:IntegrationContour}   
\end{figure}

The Dyson-Schwinger and Bethe-Salpeter equations are integral equations which are 
solved numerically. Hereby, for the presented calculation the
major technical challenge arises from the singularities of the integrands. This implies that 
one has to find integration contours in presence of cuts generated by quark propagator poles
and the pion propagator pole as well as two-pion cuts and the $\rho$ pole, the 
latter two also arising in the quark-photon vertex. To this end one has to determine,
typically by resolving
respective conditions numerically, the location of poles and cuts of the integrand. 
Then the integration contour is deformed such no cuts are crossed and no residues 
of poles are inadvertently picked up. Fig.\ \ref{fig:IntegrationContour}  shows an example
of such a contour deformation. Note that different contour deformations have to be done
for different external momenta.

\section{Results}

For the presented exploratory calculation, the parameters are adjusted in a simple manner,
especially as we aim at a qualitative understanding of the physical mechanisms involved in determining
 the shape of the pion form factor, and not so much at achieving an accurate quantitative agreement with experiment. 
As the pion-decay kernels will not only move the 
$\rho$-meson pole into the complex plane but also shift down its real value a precise 
fit of the two (effective) parameters of the Maris-Tandy model to reproduce the meson masses
would have required quite
some computational resources. In addition, only a reasonably accurate value 
for the pion decay constant, namely $f_\pi$= 138 MeV has resulted. Although this value
 is a few percent larger than the experimental one,
 $f^{expt.}_\pi=130.5$ MeV, we have adopted it in
view of a compromise in the effort needed for the parameter determination 
and the related
accuracy.
In table \ref{tab:masses} the determined meson masses as well as the pion decay
constant are shown. 

\begin{table}[ht]\caption{\label{tab:masses}
The pion mass $m_\pi$, the pion decay constant $f_\pi$, the $\rho$-meson and 
$\omega$-meson masses $m_\rho$ and $m_\omega$ for the two different used 
parameterisations of the model (labeled with $\eta$)
and for the case with gluon- and pion-exchange kernels but without decay 
kernels are shown.
The current quark mass has been set to $m_q=6.8~$MeV.
The two rightmost, separated, columns display for the case when including the decay kernels
the extracted  $\rho$-meson pole position defined as
$M_{pole}^2=M_\rho^2-iM_\rho \Gamma_\rho $.
All values for dimensionful quantities are given in GeV.}
\centering
\begin{tabular}{l|cccc||cc}
$\Lambda=0.78$&
$m_\pi$&
$f_\pi$&
$m_\rho$&
$m_\omega$&
$M_\rho$&
$\Gamma_\rho$\\
\hline
$\eta=1.5$ & 0.139 & 0.138 & 0.768 & 0.778  & 0.750 & 0.100\\
$\eta=1.6$ & 0.126 & 0.138 & 0.774 & 0.784  & 0.759 & 0.105\\
\end{tabular}
\end{table}

The pion exchange kernels lift the otherwise present degeneracy in between the iso-vector
$\rho$- and the iso-singlet $\omega$-meson. These interactions  induce a mass 
splitting $m_\omega - m_\rho = 10$ MeV 
which compares favourably with the experimental splitting of 7 - 8 MeV.

\begin{figure}[ht]
\centering
\sidecaption
\includegraphics[width=0.7\textwidth]{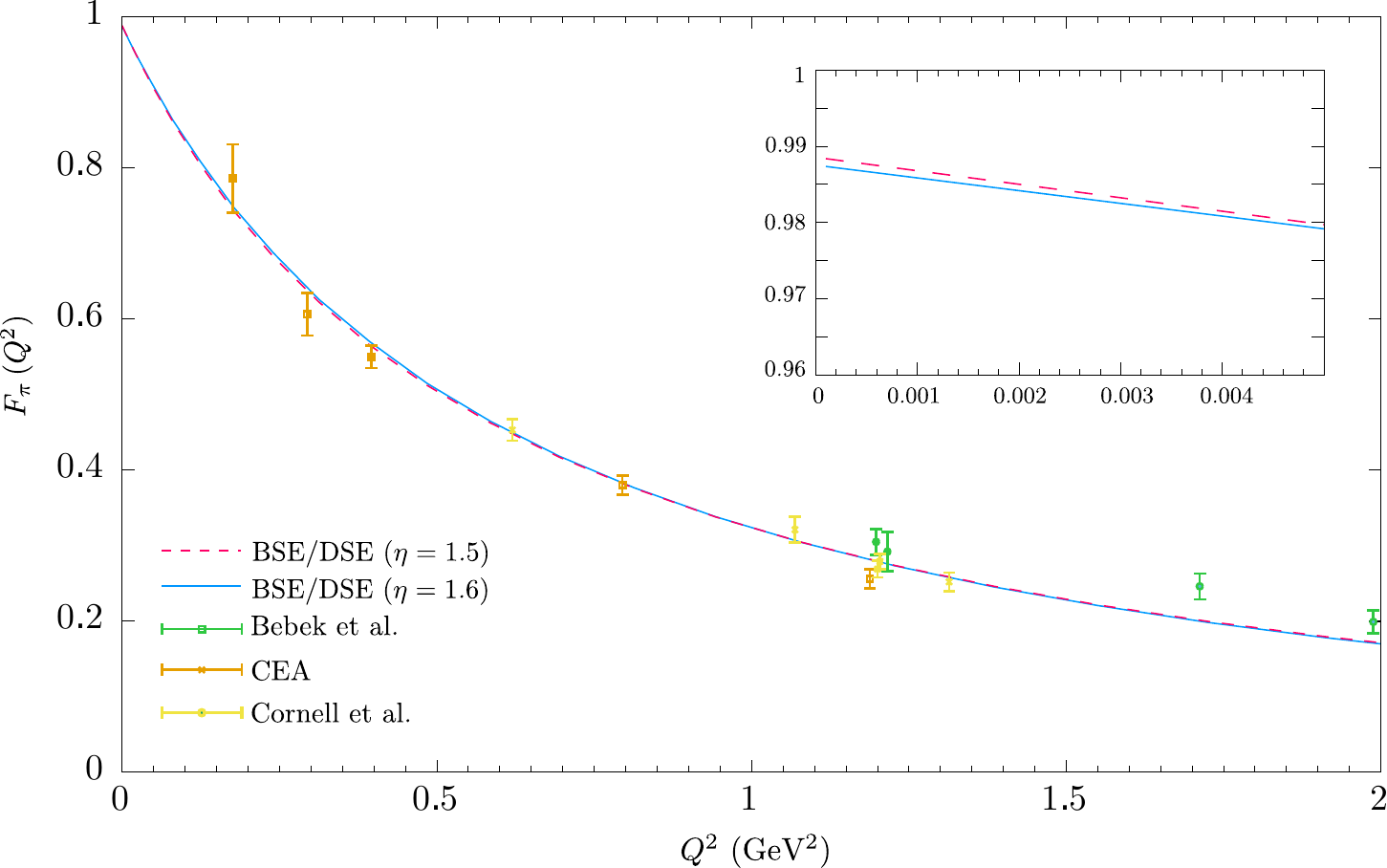}
\caption{Pion form factor in the spacelike $Q^2>0$ domain for the model parameters
 $\eta=1.5$ and $\eta=1.6$. (The inset illustrates the impact of one of the technically 
motivated approximations.)}
\label{fig:FF_spacelike}
\end{figure}

Fig.~\ref{fig:FF_spacelike} displays our results for the space-like form factor from
which we also extract the result for the pion radius, 
$\sqrt{\langle r_\pi^2 \rangle}=$ 0.68 fm.
It is reassuring that the space-like form factor, including the pion radius, shows a very 
good agreement with experimental data. As we included the most important
physical processes with effects in the time-like regime we take this as evidence that 
the imaginary part in the time-like region is
precisely enough reproduced to provide very good results for the space-like form factor.

\begin{figure*}[ht]
\caption{Absolute value of the pion form factor in the time-like $Q^2<0$ domain for  the model parameters $\eta=1.5$ and $\eta=1.6$ as described in the text and compared to experimental data \cite{Akhmetshin:2006bx}.}
\label{fig:FF_abs_timelike}
\centering
\includegraphics[width=0.99\textwidth]{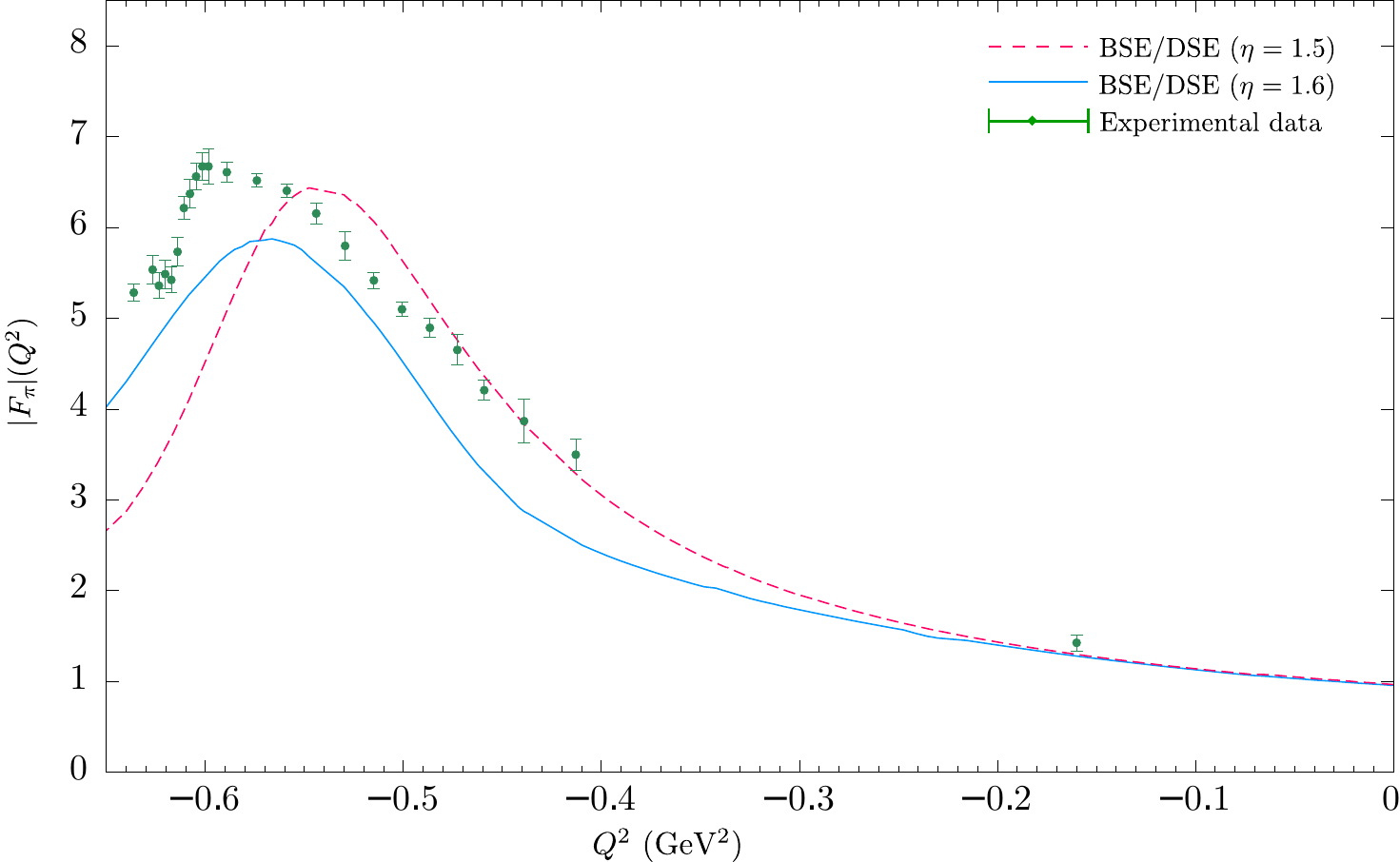}
\end{figure*}

The results for the absolute value of the
 pion form factor for time-like ($Q^2<0$) virtualities  are displayed in Figs.~\ref{fig:FF_abs_timelike}. As a consequence of the decay kernels, the pion form factor 
(as well as the quark-photon vertex)
possess a cut along the real negative axis starting at the two-pion threshold $Q^2=-4m_\pi^2$,
and therefore also an imaginary part contributes significantly to the absolute value.

With decay kernels included, the $\rho$-meson pole, present in the quark-photon vertex, 
moves into the complex plane, and consequently the pion form factor develops a bump-like
shape with approximately correct height and width. Therefore the presented calculation 
corrects a deficiency of earlier respective studies of the time-like pion form factor, 
without or even with the pion exchange term, for which the form factor diverges 
at the $Q^2$-value corresponding to the $\rho$-meson mass.

\begin{table}[ht]
\caption{\label{tab:Coeffs}
The coefficients of the rational fits to the pion form factor as discussed in the text.
All values for dimensionful quantities are given in GeV.}
\centering
\begin{tabular}{l|ll||l||l||ll||l}
& $\eta$ =1.5 & $\eta$ =1.6 & VMD expression & 
& $\eta$ =1.5 & $\eta$ =1.6 & VMD expression  \\
\hline \hline
$a_1$ & 0.5587 & 0.4149 & 0.72 & $c_1$ & 0.0591 & 0.0997 & 0\\
$a_2$ & 0.8828 & 0.6827 & 1.2  & $c_2$ & 0.1295 & 0.2383 & 0.2308\\
\hline
$b_0$ & 0.3600 & 0.3600 & 0.36 &$d_0$ & 0.3600 & 0.3600 & 0.36 \\
$b_1$ & 1.2307 & 1.2517 & 1.2 & $d_1$ & 1.1924 & 1.2464 & 1.2\\
$b_2$ & 1.0722 & 1.1000 & 1.037 & $d_2$ & 0.9973 & 1.0916 & 1.037\\
\end{tabular}
\end{table}

In order to provide a measure for the proximity of our results to the 
VMD predictions we performed
rational fits (Pad\'e fits) to the results for the real part and
for the imaginary part of the form factor and compare the coefficients
to the VMD expression without $\rho$-$\omega$ mixing. To stay as general
as possible we used first an ansatz with twelve parameters in total:
\begin{eqnarray}
Re \, F_\pi (Q^2) - F_\pi (0) & \approx &
- \frac {a_0 + a_1 Q^2 + a_2 (Q^2)^2 + a_3 (Q^2)^3}{b_0 + b_1 Q^2 + b_2 (Q^2)^2 + b_3 (Q^2)^3} \nonumber \\
Im \, F_\pi (Q^2) & \approx & \frac {c_0 + c_1 Q^2 + c_2 (Q^2)^2 + c_3 (Q^2)^3}
{d_0 + d_1 Q^2 + d_2 (Q^2)^2 + d_3 (Q^2)^3} \, . \nonumber \\
\end{eqnarray}
We obtained almost vanishing values for the coefficients 
$a_0$, $a_3$, $b_3$, $c_0$, $c_3$ and $d_3$. 
Phrased otherwise, 
 {\em the fit confirms the
structure expected from the VMD expression}. Based on this, the fits have been repeated for
\begin{eqnarray}
Re \, F_\pi (Q^2) - F_\pi (0) & \approx &
- \frac {a_1 Q^2 + a_2 (Q^2)^2 }{b_0 + b_1 Q^2 + b_2 (Q^2)^2}\, , \nonumber \\
\quad 
Im \, F_\pi (Q^2) & \approx & \frac { c_1 Q^2 + c_2 (Q^2)^2 }
{d_0 + d_1 Q^2 + d_2 (Q^2)^2 } \, .
\end{eqnarray}
The coefficients resulting from these fitting ans\"atze  as well as the ones resulting 
from VMD expression are given in table~\ref{tab:Coeffs}. 
Note that the expression based on the VMD hypothesis
is an astonishingly good representation of the results based on the microscopic model of QCD degrees of freedom: 
Other terms than VMD-predicted ones are tiny, and the elaborated calculation yields 
within error margin the VMD predicted functional form.

Another remark applies to the fact that there is no significant impact from quark propagator poles
on the pion form factor. As discussed in the Introduction this is a wanted feature in the 
view of confinement, however, it should be understood why this happens in the presented study which 
contains, despite its quite high level of sophistication,  not all QCD features.

Reassuring is nevertheless that the resulting time-like pion form factor in the kinematic region $0>Q^2>0.8$GeV$^2$ is determined by the $\rho$-meson pole and the two-pion cut
despite basing the calculation on QCD degrees of freedom.

\section{Conclusions and Outlook}

The main message of the presented study is the feasibility of studying time-like quantities 
in a Dyson-Schwinger--Bethe-Salpeter approach including intermediate resonances and thus
decay channels in the interaction kernels. Despite the applied modelling (which can be 
overcome in future investigations) and the technical limitations inherent to the approach we 
found a remarkable agreement with the experimental data.

As a side result we provided a detailed quantitative verification of the VMD model for the 
pion form factor. In view of the fact that the $\rho$-meson pole and the two-pion cut 
are determining the prominent features of the pion form factor it might be considered 
 expected that the VMD hypothesis should be, at least, qualitatively correct. 
Therefore, the more surprising,  and definitely positive, 
aspect is that a microscopic model built on QCD degrees of freedom provide 
a pion form factor shaped by the hadron substructure of the virtual photon.

One obvious way to improve the presented study is to take isospin breaking via the different
quark masses and electric charges into account. First steps in this direction have been 
undertaken \cite{Miramontes:2022mex}, however,
 a microscopic description for $\rho$-$\omega$-mixing
including the decay channels for these vector mesons, and its implementation 
into a calculation of the pion form factor, is still a task to be performed yet.

It would be also worth to extend the presented study to calculate the 
$\gamma$ $\pi$ $\pi$ $\pi$ form factor which is only non-vanishing due to the chiral anomaly.
How in such a form factor, whose soft-point value is fixed due to the anomaly and thus easily understood from the perspective of QCD, the hadron 
resonances are determining the prominent time-like features will be of general interest.

Last but not least, the question of the accessibility of the nucleons' time-like form factors arises.
The presented study paves the way for an analogous inclusion of decay kernels into 
baryon bound state equations. To this end we note that a 
functional approach based on Dyson-Schwinger and bound-state equations allows for a unified
description of mesons and baryons, see, {\it e.g.}, Ref.\ \cite{Eichmann:2016yit}, 
which quite successfully describes the hadron spectrum and space-like form factors. 
However, exactly these studies provide an estimate about the dramatic 
increase in complexity when dealing with baryon 
instead of meson bound state equations and a respective calculation of form factors based
on these equations. Nevertheless, given the increased effort from the 
experimental side and the 
resulting anticipated highly precise data over a large kinematical region certainly justifies
the correspondingly increased effort on the theoretical side.  

\bigskip \bigskip

\section*{Acknowledgements}

We are grateful to the organisers of the 
{\it International Conference on Exotic Atoms and Related Topics} (EXA21) 
for all their efforts which made eventually this (online) conference possible.\\
This work was partially supported by the the Austrian Science Fund (FWF) under project number P29216-N36.\\
 A.S.\ Miramontes acknowledges CONACyT for financial support.\\
The numerical computations have been performed at the high-performance compute cluster of the University of Graz.

\end{document}